\documentstyle[preprint,revtex]{aps}

\tightenlines
\begin{document}

\begin{title} Magnetic field effects on non-periodic superlattice
structures
\end{title}

\author{H. Cruz$^{\dag}$, F. Piazza, and L. Pavesi$^*$}
\begin{instit} Dipartimento di Fisica. Universit\`{a} di Trento.
I-38050 Povo (Trento). Italy
\end{instit}

\begin{abstract}

\noindent A simple numerical method to study the effect of an applied
magnetic field on the energy spectrum of non-periodic superlattice
structures is presented. The magnetic field could be either parallel or
perpendicular to the growth direction. Our method is based on the transfer
matrix technique and on the effective mass approximation. We discuss the
advantages and disadvantages of the proposed approach using several
examples. In particular, we study the perturbation to the energy spectrum
of periodic superlattice induced by the introduction of an enlarged well.
We found that these perturbations are negligible for B//z but relevant for
B$\perp $z. Preliminary results for Fibonacci superlattices in magnetic
fields are presented as well. In these quasi-periodic structures the energy
levels become strongly dispersive in presence of a perpendicular magnetic
field.

\end{abstract}
\noindent to be published in Semiconductor Science and Technology
\section{INTRODUCTION}
\label{sec:level1}

In 1970, Tsu and Esaki proposed superlattices as new structures with
peculiar properties (minibands) which allowed new physical phenomena to be
studied.\cite{esaki} In recent years, the improvement of epitaxial growth
techniques renders disposable samples to study the properties of transport
in the vertical direction (i. e. parallel to the growth direction, $z$).
Resonant quantum tunneling through double barrier heterostructures,
transport along superlattice minibands, and optical studies in biased
multiquantum well structures have recently attracted great
attention.\cite{capasso} In short period superlattices, the coupling
between adjacent wells leads to transport through superlattice miniband
states.\cite{deveaud} In addition, the present availability of high
magnetic fields stimulated many experimental studies of
magnetotransport.\cite{Kopev} A problem of increasing interest is the
effect of a transverse magnetic field, B, on the tunneling through a
barrier separating either two semiconductors\cite{snell} or two
superlattices \cite{davies}. Magnetotransport in transverse fields has also
been studied in a semiclassical macroscopic framework \cite{movagar}. Other
topics concern the effect of an applied magnetic field perpendicular to the
semiconductor layers of an heterostructure \cite{silva,tejedor}. In this
case, the magnetic quantization in the layer plane collapses the
two-dimensional (2D) electron gas in a discrete set of ciclotronic orbits
of zero degrees of freedom (0D system).

The problem of the influence of an external magnetic field on the
superlattice energy levels have been throughout studied by theoretical
techniques. Two different approaches have been used: the first is based on
a numerical integration either of the Schr\"{o}dinger equation\cite{maan}
or of a reduced Luttinger Hamiltonian\cite{fasolino} and the second solves
the Schr\"{o}dinger equation using the method of expansion of the solutions
in series of analytical functions (like sine functions\cite{Xia1989},
ipergeometrical confluent functions\cite{Cruz} or parabolic cylinder
functions\cite{Huang}). In this paper we propose a different approach
based on the
extension of the well-known Transfer Matrix (TM) method\cite{Kane,pavesi2}
to
superlattices in magnetic field. The interest in our approach is the
possibility to study superlattices formed by whatever a succession of
layers. The TM method has been widely used to solve
the problem of non-periodic structures.\cite{Erdos} Its simplicity is
related to the use of the product of only 2x2 matrices. This renders the
method very appealing because it is easily implementable on small
computers.

In the Appendix we present the details of our approach to the Transfer
Matrix method in absence of magnetic field. For simplicity we have
restricted our discussion to the electronic problem, i. e. we neglect the
holes. In section II we study the case of $B||z$, where $z$ is the growth
direction: the superlattice miniband width is varied due to the magnetic
quantization in the heterostructure layer planes, i.e., to the conservation
of the Landau level index along the different layers. In section III, we
analyzed the case of $B{\perp}z$. A new equation which couples the motion
in the layer plane and in the vertical direction should be solved to obtain
the energy levels. Then, we study two particular problems: the enlarged
well in a superlattice (section IV) due to its importance for vertical
transport studies and the Fibonacci superlattice (section VI) as an example
of non-periodic superlattice. In section V we have computed the density of
states for magnetic fields oriented either parallel or perpendicular to z.
Finally, section VII ends the paper with a summary of the main results.

\section{MAGNETIC FIELD PARALLEL TO z}
\label{sec:level3}

Through the effective-mass approximation it is possible to separate the
electronic wave-function in a component (usually assumed to be a plane
wave) for the motion in the layer plane and in a component (solution of the
equation (\ref{1})) for the motion in the $z$ direction. A magnetic field
applied in the $z$ direction does not introduce a mixing of the parallel
and the perpendicular electron wave functions. However, it changes the
plane waves for the motion in the layer plane into harmonic oscillator wave
functions and it collapses the free electron dispersion relation in a
discrete set of Landau levels.\cite{tejedor} In this case, the eigenvalues
$E_{n}$ are

\begin{equation}\label{9}
E_{n}=E_{z}+{\hbar}{\omega}_{w}(n+1/2)
\end{equation}

\noindent with $n=0,1,2,..$ the Landau level number, $E_{z}$ the
eigenvalue for the motion in the vertical direction, $\omega _w=eB/m^{*}c$
the cyclotron frequency for the GaAs effective-mass $m^*$, $e$ the
electronic charge and $c$ the light velocity in vacuum. In (\ref{9}), we
have neglected the spin magnetic energy.\cite{Bimberg}

To obtain $E_{z}$, it is necessary to solve the effective-mass
Schr\"{o}dinger equation

\begin{equation}
\left[ - \frac{{\hbar}^{2}}{2m^{*}}{\nabla}^{2}
+V_{n}(B,z) \right] {\psi}(z)=E_{z}{\psi}(z)
\end{equation}

\noindent where the magnetic field is included by changing the
ordinary superlattice potential, $V_{SL}(z)$, with an effective
one-dimensional potential, $V_{n}(B,z)$, different for each Landau
level $n$,

\begin{equation}\label{11}
V_{n}(B,z)=V_{SL}(z)+ {\hbar}(n+1/2)[{\omega}_{w}-
{\omega}_{b}]=V_{o}+ {\hbar}(n+1/2)[{\omega}_{w}-
{\omega}_{b}]
\end{equation}

\noindent in the Al$_{x}$Ga$_{1-x}$As barriers and

\begin{equation}\label{12}
V_{n}(B,z)=V_{SL}(z)=0
\end{equation}

\noindent in the GaAs wells. $V_o$ is the conduction band discontinuity
between GaAs and Al$_{x}$Ga$_{1-x}$As, and ${\omega}_{b}$ is the cyclotron
frequency for the Al$_{x}$Ga$_{1-x}$As effective-mass. The potential
described in equations (\ref{11}) and (\ref{12}) reflects the conservation
of the Landau level number through the layers.\cite{silva} Consequently,
$E_{z}$
depends on the applied magnetic field.

Figure 1 shows the highest and lowest electron energy level of the
miniband of a
30/30 superlattice (30{\AA} thick wells and barriers, with $x=0.33$ in the
barriers) versus the
applied magnetic field for different values of $n$.\cite{pa:ra}
By increasing the applied magnetic field, the electron energy
levels move to low energy. The energy shift is larger for higher Landau
level number. This effect is explained by equation (\ref{11}). The
Al$_{x}$Ga$_{1-x}$As effective mass is higher than the GaAs effective mass,
so that the cyclotron frequency in the Al$_{x}$Ga$_{1-x}$As layers is lower
than in the GaAs layers. Thus, the second part on the right hand side of
equation (\ref{11}) is a negative term proportional to $B$ and $n$. If $B$
or $n$ increases, the potential barrier is lowered, and the electron energy
levels are shifted to lower energies. Hence, the coupling of the electron
wave functions in adjacent wells is augmented increasing the miniband
width. To evaluate this effect, we compare the dependence
of $E_z$ on $B$ for various $n$ with the corresponding ciclotronic
kinetic energy ${\hbar}{\omega}_{w}(n+1/2)$. For $B$ measured in Tesla,
${\hbar}{\omega}_{w}=1.72\; B$ meV. So that the conservation of the quantum
number $n$ changes $E_{n}$ by 5 \%. This effect could be increased if a
higher aluminum concentration is used.

\section{MAGNETIC FIELD PERPENDICULAR TO z}
\label{sec:level4}

In this case, the Hamiltonian describes a spinless particle in a layer of
the superlattice, with an effective mass $m^{*}$ and charge $e$, subject to
a constant and uniform transverse magnetic field {\bf B}, i.e.

\begin{equation} H=\frac{1}{2m^{*}}({\bf p}- \frac{e}{c} {\bf
A})^{2}+V_{SL}(z)
\end{equation}

\noindent where ${\bf p}=-i \hbar \nabla $, and {\bf A} is the magnetic
vector potential. We fix the direction $B||x$ to simplify the equations.
The Hamiltonian takes the form
\cite{tejedor}

\begin{equation} H=- \frac{\hbar^2}{2m^{*}} \nabla^{2}+
\frac{e^{2}B^{2}z^{2}}{2m^{*}c^{2}} -\frac {ie{\hbar}Bz}{m^{*}c}
\frac{\partial}{\partial y}+V_{SL}(z)
\end{equation}

\noindent where the {\it gauge} ${\bf A}=(0,-Bz,0)$ has been used. By using
a plane wave in the {\it x} and {\it y} direction and neglecting spin
effects, it is possible to find the envelope wave-function $\psi (z)$ that
describes the motion of the carriers along the {\it z} direction of the
heterostructure through the equation

\begin{equation} H \psi (z)=E \psi (z)
\end{equation}

\noindent with $E$ the electron energy and the effective mass Hamiltonian,
$H$, given by

\begin{equation}\label{16}
 H =- \frac{\hbar^{2}}{2m^{*}} \frac{d^{2}}{dz^{2}}+
\frac{1}{2} m^{*} \omega_w^{2} (z-z_{0})^{2}+V_{SL}(z) =
- \frac{\hbar^{2}}{2m^{*}} \frac{d^{2}}{dz^{2}}+ V_{eff}(z)
\end{equation}

\noindent where the definitions $z_{0}=- \hbar k_{y}c/eB$ have been used.
$k_y$ is the wave-vector in the y-direction. Note that the eigenvalues E
depend on $z_{0}$.

To calculate the electron energy eigenvalues for different $z_{0}$
(dispersion relation) we have assumed a flat band approximation for the
potential $V_{eff}(z)$. This consists in taking $V_{eff}(z)$ constant in
each layer. For a layer {\it i} between two interfaces $z_{j}$ and
$z_{j+1}$, we chose a mean value for the magnetic potential. We define
$z_{i}=(z_{j}+z_{j+1})/2$ and the second term on the right hand side of
equation (\ref{16}) yields a constant value
$V_{i}=m^{*}\omega^{2}(z_{i}-z_{0}) ^{2}/2$. Then, in order to use the
transfer matrix method, we substitute $E{\rightarrow}E-V_{i}$ in each
formula of the Appendix. This approximation is based on the large number of
barriers and wells that forms a superlattice. The averaged flat band
potential is expected to be a good approximation of the real potential
$V_{eff}(z)$ if the wave function extends over many wells and barriers. In
this case, we can neglect the parabolic shape of the potential $V_{eff}(z)$
in each individual well and barrier. Increasing the number of layers, the
accuracy of the averaged flat band approximation is increased. In this way,
the application of a parabolic magnetic field on a superlattice is
modelized by means of a variable height flat band potential. This method is
similar to the approximations that R. Tsu and L. Esaki have used to compute
the effect of an external electric field on a superlattice.\cite{tsu}

Our approximation has a low influence for low magnetic fields. In addition,
we have found a very small numerical differences for the dispersion
relations of electrons in superlattice under high magnetic fields (up to
B=20T) with respect to the results of J.C. Maan \cite{maan}. In Ref.
\cite{maan} the Schr\"odinger equation (\ref{16}) has been numerically
integrated. Moreover, to check further the approximation used in our
calculations we have tried to reproduce more accurately the potential
$V_{eff}(z)$ in each layers. In a barrier or well placed between the
$z_{j}$ and $z_{j+1}$ interfaces we discretize the layer in a finite number
($L$) of regions. The interfaces between the new $L$ regions are found at
$z_{\ell}=z_{j}+(z_{j+1}-z_{j}) \ell/L$. We define a flat band potential in
each $\ell$ region as $V_{\ell}=V_{SL}(z_\ell) +
m^{*}\omega_w^{2}(z_{\ell}-z_{0})^{2}/2$. Increasing the number $L$, the
new step-like potential will be closer and closer to $V_{eff}(z)$. Table I
reports the electron energy levels in a 30/30 Al$_{0.33}$Ga$_{0.67}$As-GaAs
superlattice with $B=5T$ for different $L$ values and a fixed $z_{0}$.
Including a large number of quantum wells in the superlattice (20 wells)
the energy levels obtained with $L=2,3,...$ are slightly different from the
results with $L=1$. The $L=1$ case represents a very good compromise
between computational efficiency and the accuracy of the results. For $L
\neq 1$ the number of the matrix products are changed from $N-1$ (N is the
number of layers) to $(N-1)L$.

We have plotted in Fig. 2 and 3 the calculated electron dispersion
relations ($z_{0}$ versus $E$) using 20 wells and $L=1$ for a 30/30 and
60/60 superlattices at $B=5$T and $4$T, respectively. These are
representative of two electronic dispersion relations with one miniband
(30/30 superlattice) or two minibands (60/60 superlattice). The dispersion
relation is flat if the eigenvalues belong to the minibands calculated in
absence of magnetic field and it is almost parabolic if not.\cite{maan}
Anti-crossing in the energy levels occurs for several $z_o$ values. For the
60/60 superlattice, it is found that the anticrossing occurs also between
the energy levels of different minibands. From a semiclassical point of
view, the anticrossings between both minibands correspond to cyclotron
orbits of electron states localized in different minibands. In this case,
one electron could tunnel from one to the other superlattice minibands.

In the literature, other methods based on the transfer matrix method have
been proposed to study the effect of the transverse magnetic field on the
superlattice energy spectrum (see e. g. \cite{Cruz,Huang}). However, these
other methods use some analytical solutions of equation (\ref{16}) in each
layer (ipergeometrical confluent functions\cite{Cruz} or parabolic cylinder
functions\cite{Huang}). The numerical calculation of these functions
requires long CPU times and introduces numerical errors and
instabilities that in our scheme are avoided. Using the analytical method
of Ref. \cite{Cruz} to calculate the energy levels for a double barrier
structure one needs 10 times longer CPU time than to compute the same
structure with the present method with $L=6$.

\section{THE ENLARGED WELL PROBLEM IN SUPERLATTICES}
\label{sec:level5}

Recently, Chomette {\it et al} \cite{chomete} have proposed to use an
enlarged well (EW) as a marker of the arrival of carriers at a certain
distance from the surfaces. This method allows to study vertical transport
in superlattices by means of optical techniques. Including a well in the
superlattice, which is wider than the superlattice wells, introduces a
localized state in the energy spectrum. The level due to the EW has a lower
energy than the superlattice miniband. In this way, the photoluminescence
from the EW can be used as a measure of the transport through the
superlattice.\cite{chomete} In absence of any applied magnetic field, it is
found that the original superlattice miniband is slightly affected by the
introduction of an EW, so that, the results obtained with this technique
can be used to understand the transport through superlattice minibands
without an EW. The enlarged well technique has been used to study
magnetotransport in superlattice in the cases of $B||z$\cite{piazza} and
$B{\perp}z$.\cite{Kopev,christianen} In this section, we analyze the
perturbation of the superlattice energy spectrum due to the EW when an
external magnetic field parallel and perpendicular to the growth direction
is applied.

For parallel magnetic field, the electron energy levels are easily found
through the method of calculation described in section II. Table II
reports, for B=50 T and n=0, the electronic energy levels for a 30/30
Al$_{0.33}$Ga$_{0.67}$As-GaAs superlattice and for the same superlattice
with an EW of 60{\AA} in the middle. The electronic energy levels are only
slightly changed by the presence of the EW even for this very high B value.
Consequently, it is possible to use an EW as a marker to study
magnetotransport in superlattices with $B||z$ by means of optical
techniques.

For $B{\perp}z$ and using the method of section III, we show in Fig. 4 and
5 the energy levels for the same superlattices as those used to obtain Fig.
2 and 3 except that a 60{\AA} and 120{\AA} EW are introduced in the middle
of the 30/30 and 60/60 superlattice, respectively. The introduction of the
EW breaks the periodic symmetry of the superlattice and introduces a weaker
inversion symmetry respect to the center of the EW. The interaction between
the unperturbed dispersion relation of the superlattice and of the EW
results in a situation very different from that shown in Fig 2 and 3. In
particular, the flat dispersion reminiscent of the electronic miniband
becomes dispersive after the introduction of the EW. This is due to the
anticrossing between the EW states and the SL states. This perturbation of
the SL energy levels disappears far away from the EW region. In a
semiclassical point of view, the anticrossing between the EW energy level
and the superlattice miniband corresponds to skipping orbits centered in
the barriers on the left and right sides of the EW.\cite{snell} One
electron in a superlattice state can tunnel to the EW localized state. As a
consequence of this interaction, for $B{\perp}z$, the EW effect on the
superlattice energy spectrum should be considered when the EW technique is
used to study magnetotransport through a superlattice.

\section{DENSITY OF STATES}
\label{sec:label6}

In quantum wells, the two dimensional density of states, $\rho (E)$, is
usually taken as a step function.\cite{libro} In a superlattice composed by
a finite number of layers, $\rho (E)$ is a series of step, one for each
level in the miniband. A parallel magnetic field collapses the 2D energy
levels in a discrete set of 0D Landau states. In this case, the step like
function for $\rho (E)$ is changed to a discrete set of Dirac
$\delta $-functions.\cite{libro} Including a Gaussian broadening of
the $\delta $-functions caused from the presence of disorder, $\rho (E)$
results

\begin{equation}\label{17}
{\rho}(E)=C{\sum}_{m}{\sum}_{n}exp \{
-[E_{m}+{\hbar}{\omega}_{w} (n+1/2)-E]^{2}/{\sigma}^{2} \}
\end{equation}

\noindent where $C$ is a normalization factor, $E_{m}$ are the energy
levels of the superlattice in the {\it z} direction with the index $m$
running over the superlattice miniband, $n$ is the landau level index and
$\sigma$ is the gaussian width. We have neglected the second term on the
right hand side of equation (\ref{11}) to obtain the energy eigenvalues
$E_{m}$, i.e. to compute equation (\ref{17}) we have used the $E_{m}$
values obtained in absence of a magnetic field. Figure 6 reports $\rho(E)$
for a 30/30 superlattice for different magnetic fields. The Landau
quantization changes $\rho(E)$ in a set of isolated peaks. For each energy
level in the miniband, the distance between two Landau states is given by
the cyclotron frequency, $\hbar {\omega_{w}}$. Increasing the applied
magnetic field, the cyclotron frequency, which is proportional to B,
increases, and then, the distance between different peaks.

For B$\perp $z, the density of states is,

\begin{equation}\label{18}
\rho(E)= \frac{eBL_{x}}{c} \frac{1}{(2 \pi \hbar)^{2}}
\sum_{m} \int_{0}^{L_{z}}
\sqrt{ \frac{m^{*}}{2[E- E_{m}(z_{0})]}} \theta [E-E_{n}(z_{0})]dz_{0}
\end{equation}

\noindent where $E_{m}(z_{0})$ are the solutions of equation (\ref{6})
which depend on the cyclotron orbit center $z_0$, and $L_{x}$, $L_{z}$ are
the superlattice dimension. The square root dependence of $\rho $(E) is due
to the 1D motion in the $x$-direction which is affected neither by the
superlattice potential nor by the magnetic field. A singularity in the
denominator of the square root occurs at $E=E_{m}(z_{0})$ and it leads to
important modifications in the optical and transport properties of these
systems.

Figure 7 shows different $\rho $(E) computed using eq. (\ref{18}). For a
60/60 superlattice at B=4 T (full line in Fig. 7), the existence of two
electronic minibands is observed at about 60 meV and 220 meV. The second
electronic minibands is wider than the first one and this allows the
resolution of distinct peaks in $\rho $(E). These peaks show a
characteristic 1/$\sqrt {E}$ dependency. In Fig. 7, we compare also the
$\rho $(E) results for B=5T of a 30/30 superlattice in presence (solid
line) and in absence (dashed line) of a 60 {\AA } EW. In the miniband
energy region ($\simeq $150 meV), the effect of the EW induces a loosing of
the 1D typical divergence, especially for the states near the high energy
edge. The direct contribution of two non-dispersive EW eigenvalues in the
density of states is marked by two arrows. If the B value is increased no
more non-dispersive levels exist and the 1D divergence in the miniband
region will disappears.

\section{THE FIBONACCI SUPERLATTICES}\label{sec:level8}

Another interesting problem which can be tackled by the present approach
is quasi-periodic superlattices. These
structures do not exhibit translational invariance but long range order.
The interest stems from the fact that the Bloch theorem is inapplicable and
that they provide an intermediate class between periodic superlattice with
extended electronic states and randomly disordered superlattice with
exponentially localized electronic states. Merlin et al.
\cite{me:ba} proposed to grow quasi periodic superlattice using the
Fibonacci sequence. These superlattices, called Fibonacci superlattices
(FSL), are interesting because using a simple recurrent law one can obtain
series of increasing period starting from usual simple superlattice and
ending
with a completely non periodic superlattice.\cite{me:ie,la:et,da:xi}
We recall that a
quasi-periodic FSL of order $f $ is constructed by $f $ applications
of the Fibonacci transformation law $A \rightarrow AB$ and $B \rightarrow
A$ starting from $A$, yielding the following sequences: B ($f = -1$), A
($f = 0$), AB ($f = 1$), ABA ($f = 2$), ABAAB ($f = 3$), ... .

In Fig. 8 we show the electronic energy dispersion for different FSL and
for B = 0 and B = 5 T. The evolution with the magnetic field strength of
the energy level scheme is shown in Fig. 9 for a FSL with index f=1 and for
B = 0, 1, 3, 5 T. In our calculations, following Ref. \cite{la:et,pavesi2},
the block A was formed by 28 {\AA } of $GaAs$ and 30.8 {\AA } monolayers of
Al$_{0.23}$Ga$_{0.77}$As, and the block B by 28 {\AA } monolayers of $GaAs$
and 67.2 monolayers of Al$_{0.23}$Ga$_{0.77}$As. We have repeated the
Fibonacci series to obtain a total superlattice thickness of $\approx 1000
\AA$ for the different FSL but the f=5 FSL which has a superlattice
thickness of 1500 \AA .

For B=0, our method yields the same results as the calculations of Ref.
\cite{la:et,pavesi2}. In summary, two kinds of states are formed in the
spectrum of a FSL. The first is formed by single-A pattern (typical of $f =
1$ FSL), the second kind is formed by double-A pattern (typical of $f = 2$
FSL). A double-A pattern is analogous of two coupled QW separated by a
thick barrier. Little overlap exists between these two kinds of states for
the $f = 1$ and the $f = 2$ FSL, whereas the electronic energy spectrum of
the other FSL presents overlaps with either the $f = 1$ FSL energy bands or
$f = 2$ FSL energy bands (see in Fig. 8 the B=0 T results for f=3 or f=5).
The effect of an external magnetic field is the breaking of this simple
scheme. In particular, for $f$=0 (periodic superlattice)
and B= 5 T we get non-dispersive states in
the region of the superlattice miniband and dispersive states for higher
energies. While for $f \neq $ 0, almost all the states are dispersive. The
quasi-periodicity of the FSL is lost and the energy dispersions are very
similar for different $f$-values. The magnetic field localizes the
electronic orbit in finite spatial regions. Hence the electron are subject
to the local potential ($V_{eff}$). $V_{eff}$ does not show the quasi
periodicity of the Fibonacci superlattice
potential. This fact is shown in Fig. 9 for $f$=1. By
decreasing the electron cyclotronic orbit (i. e. increasing B) increases
the dispersion of the states. For B=1 T the energy dispersion is flat.
Increasing the magnetic field the energy dispersion changes: for B=3 T we
observe both dispersive and non dispersive states while for B=5 T almost
all the states are dispersive.
We note
that the band of states
at $\approx  $ 250-300 meV which develops for B= 5 T and for f=0-3 in Fig.
8
is due to the finite size of the superlattices.

\section{SUMMARY}
\label{sec:label7}

In summary, we have presented a simple method to compute the effect of an
external magnetic field on the energy spectrum of superlattices. This
method is based on the transfer matrix approach. Due to the strong
non-parabolicity and the valence band mixing of hole states in
superlattices, we have applied our approach only to the electronic states
in this paper.

A magnetic field parallel to $z$ leads to a full quantization of the
electronic motion in the layer plane. The conservation of the Landau level
indexes along the superlattice increases the energy miniband width and
induces a dependence of the energy levels on the Landau index and the
applied magnetic field value. The density of states is changed from a
step-like function to a set of $\delta $-like functions.

For $B{\perp}z$ case, we have computed the dispersion relations and the
density of states for electrons in a superlattice. The energy spectrum
shows the existence of energy regions with dispersive and non dispersive
energy states. The density of states is characterized by a typical 1/$\sqrt
{E}$ dependency for non-dispersive states and by a smooth increase for
dispersive states.

We have applied our method to study the enlarged well problem in
superlattices with an applied magnetic field. With $B||z$, we have found
that the presence of the EW is only a small perturbation on the
superlattice energy spectrum which permits the use an EW in the study of
magnetotransport through superlattices. For $B{\perp}z$ case, the EW
strongly perturbs the superlattice energy spectrum.

Fibonacci superlattices which are a particular class of non-periodic
structures show some interesting aspects due to the loss of the
quasi-periodicity in magnetic fields. Our preliminary results suggest that
due to the increase of the magnetic localization of the electrons the
self-similar energy spectrum typical of these quasi-periodic structure is
lost. A magnetic field of 5 T is enough to render dispersive all the
electronic states
while for periodic superlattice the energy dispersion is still
flat at this value of the magnetic fied.

\acknowledgments

H.C. wishes to acknowledge the financial support of {\it Gobierno
Aut\'onomo de Canarias.}

\unletteredappendix{TRANSFER MATRIX METHOD}
\label{sec:level2}

Through the effective-mass approach it is possible to separate the wave
function in a component parallel to the layers and a component in the
growth direction, $z$. The eigenvalues in the $z$ direction, $E_{z}$, and
the envelope wave functions, $\psi (z)$, in a superlattice can be
calculated using spherical masses and parabolic dispersions in the layer
plane.\cite{andreani,graft} In this way, the motion of electrons and holes
in the growth direction can be decoupled and reduced to the solution of a
one-dimensional Schr\"{o}dinger equation for each kind of particle. Let us
assume to solve the problem of electrons. The Schr\"{o}dinger equation is:

\begin{equation}\label{1}
\left[ -\frac{\hbar^{2}}{2m^{*}(z)}\nabla^{2}+V_{SL}(z)
\right] \psi
(z)=E \psi (z)
\end{equation}

\noindent where $m^{*}$ is the effective mass for electrons in the
different semiconductor layers and $V_{SL}(z)$ is the superlattice
potential. The superlattice potential is given by the difference of the
energy gap between different layers which we have approximated using a one
dimensional {\it Kronig-Penney}-like potential.\cite{kronig} The energy
zero is defined at the bottom of the GaAs conduction band for the
electrons. The envelope wave function solution of (\ref{1}) in the {\it
i}th well is

\begin{equation}\label{2}
\psi_{i}(z)=A_{i}cos(k_{i}z)+B_{i}\frac{m_{i}^{*}}{k_{i}}sin(k_{i}z)
\end{equation}

\noindent where $A_{i}$ and $B_{i}$ are two constants,
$k_{i}=(2m_{i}^{*}E)^{1/2}/\hbar$ and $m_{i}^{*}$ is the effective mass in
the {\it i}th layer. The envelope wave function in a {\it i}th barrier is

\begin{equation}\label{3}
\psi_{i} (z)=\frac{A_{i}}{\sqrt{2}}e^{-\kappa_{i}z}+\frac{B_{i}}{\sqrt{2}}
\frac{m_{i}^{*}}{\kappa_{i}}e^{\kappa_{i}z}
\end{equation}

\noindent where $\kappa_{i}=\sqrt{2m_{i}^{*}(V_{0}-E)}/\hbar$, with $V_{0}$
as the height of the potential barrier.

Using a carrier current conserving boundary condition, the continuity of
$\psi (z)$ and $\psi '(z)/m^{*}$, the coefficients $A_{i}$ and $B_{i}$ of
the layer {\it i} are joined to those of the layer {\it i}+1 by a transfer
2x2 matrix ${\cal S}_{i \rightarrow i+1}$ of determinant 1 defined in the
following way \cite{pavesi2}

\begin{equation}\label{4}
\left(\begin{array}{c} A_{i+1} \\ B_{i+1} \end{array}\right) ={\cal S}_{i
\rightarrow i+1}
\left(\begin{array}{c} A_{i} \\ B_{i} \end{array}\right)
\end{equation}

\noindent ${\cal S}_{i \rightarrow i+1}$ is equal to
$(C_{i+1}^{w})^{-1}C_{i}^{b}$ if the layer {\it i} is a barrier and the
({\it i+1})th layer a well, and to $(C_{i+1}^{b})^{-1}C_{i}^{w}$ if the
{\it i}th layer is a well and the ({\it i+1})th layer a barrier. Both
functions $C$ are calculated at the interface between the {\it i} and ({\it
i+1}) layer. $C_{i}^{w}$ is defined as

\begin{equation}\label{5}
C_{i}^{w}= \left(\begin{array}{cc} cos(k_{i}z) &
\frac{m_{i}^{*}}{k_{i}}sin(k_{i}z) \\
\frac{-k_{i}}{m_{i}^{*}}sin(k_{i}z) & cos(k_{i}z) \end{array}\right)
\end{equation}

\noindent and $C_{i}^{b}$ as

\begin{equation}\label{6}
C_{i}^{b}=\left(\begin{array}{cc} e^{-\kappa_{i}z} &
\frac{m_{i}^{*}}{\kappa_{i}}e^{\kappa_{i}z}
\\
\frac{-\kappa_{i}}{m_{i}^{*}}e^{-\kappa_{i}z} & e^{\kappa_{i}z}
\end{array}\right)
\end{equation}

By consecutive applications of the matrix ${\cal S}_{i \rightarrow i+1}$ we
form the matrix ${\cal M}$ which joins the first and last layers of the
superlattice

\begin{equation}\label{7}
\left(\begin{array}{c} A_{N} \\ B_{N} \end{array}\right)={\cal M}
\left(\begin{array}{c} A_{1} \\ B_{1} \end{array}\right)=
\prod_{i=N}^{2}{\cal S}_{i-1 \rightarrow i}
\left(\begin{array}{c} A_{1} \\ B_{1} \end{array}\right)
\end{equation}

\noindent with $N$ the number of superlattice layers. Imposing vanishing
wavefunctions in the first and last superlattice layers, we find that
$A_{1}=B_{N}=0$, and from (\ref{7}) the (2,2) element of the matrix ${\cal
M}$
results in

\begin{equation}\label{8}
{\cal M}_{22}=0
\end{equation}

\noindent The solutions of the eigenvalue equation (\ref{8}) yield the
miniband
spectrum for electrons in the superlattice.

This method, which use the product of 2x2 matrices, can be easily
implemented on small computers and it permits to calculate any finite
succession of layers and not only periodic superlattices. On the other
hand,
we neglect energy band non-parabolicity for the electrons and the
heavy-hole light-hole mixing in the valence band. Consequently a big error
is expected when these effects are important. To estimate the errors on our
calculations, we have compared our results with the ones of
Ref.\cite{libro} where energy band non-parabolicity and valence band mixing
have been considered. The differences between the results for short period
superlattices (period shorter than 120 \AA) are of about 2\% for
electron energy levels and of about 10\% for hole energy levels. Due
to the neglection of valence band mixing, we have found a worst agreement
for the heavy hole energies than for electron energies, where the agreement
is rather satisfying. To go beyond our approximations, larger matrices are
necessary,\cite{Ram} which result in a larger computational effort.

\figure{Lowest and highest electronic energy level in the miniband
 of a 30/30 (30 {\AA} wide wells
and barriers)
Al$_{0.33}$Ga$_{0.67}$As-GaAs superlattice versus the applied parallel
magnetic field B for different Landau level index. Solid line: n=0,
Dotted line: n=1, Dashed line: n=2. The zero of the energy
scale is the bottom of the GaAs conduction band.}

\figure{Solid line: electron dispersion relation for the 30/30 (30 {\AA}
wide wells and barriers) Al$_{0.33}$Ga$_{0.67}$As-GaAs superlattice for a
perpendicular magnetic field B=5T. Dashed line: position of the equivalent
superlattice potential. The zero of the energy scale is the bottom of the
GaAs conduction band.}

\figure{Solid line: electron dispersion relation for the 60/60 (60 {\AA}
wide wells and barriers) Al$_{0.33}$Ga$_{0.67}$As-GaAs superlattice for a
perpendicular magnetic field B=4T. Dashed line: position of the equivalent
superlattice potential. The zero of the energy scale is the bottom of the
GaAs conduction band.}

\figure{Solid line: electron dispersion relation for the 30/30 (30 {\AA}
wide wells and barriers) Al$_{0.33}$Ga$_{0.67}$As-GaAs superlattice with a
60{\AA} enlarged well in the middle,
for a perpendicular magnetic field B=4T. Dashed
line: position of the equivalent superlattice potential. The zero of the
energy scale is the bottom of the GaAs conduction band.}

\figure{Solid line: electron dispersion relation for the 60/60 (60 {\AA}
wide wells and barriers) Al$_{0.33}$Ga$_{0.67}$As-GaAs superlattice with a
120{\AA} enlarged well in the middle,
for a perpendicular magnetic field B=6T. Dashed
line: position of the equivalent superlattice potential. The zero of the
energy scale is the bottom of the GaAs conduction band.}

\figure{Density of states in arbitrary units versus energy for a 30/30 (30
{\AA} wide wells and barriers) Al$_{0.33}$Ga$_{0.67}$As-GaAs superlattice
in a parallel magnetic field. Solid line: B=1T. Dashed line: B=2.5T.The
zero of the energy scale is the bottom of the GaAs conduction band.}

\figure{Density of states versus energy for a 60/60 (60 {\AA} wide wells
and barriers) Al$_{0.33}$Ga$_{0.67}$As-GaAs superlattice (solid line) at
B=4T. The density of states for a 30/30 superlattice is also shown for B=5T
without (solid line) and with (dashed line) an enlarged well in the middle.
The magnetic field is perpendicular to the growth axis. The zero of the
energy scale is the bottom of the GaAs conduction band. The arrows
indicate the enlarged well bound states.}

\figure{Electronic energy levels of different Fibonacci superlattices with
index $f$ as a function of the cyclotronic orbit center ($z_o$) for a
perpendicular magnetic field B= 0 T (discs) and B=5 T (points). For a
definition of the Fibonacci index $f$ and for the superlattice parameters
see
the text. For B= 0 T the energy levels do not depend on $z_o$. The energy
barrier in the superlattice is 177 meV and the zero of the energy scale is
the bottom of the GaAs conduction band.}

\figure{Electronic energy levels of a Fibonacci superlattice with index $f$=1
as a function of the cyclotronic orbit center $z_o$ and for different
values of the perpendicular magnetic field B. For B= 0 T the energy levels
do not depend on $z_o$. For a definition of the Fibonacci index $f$ and for
the superlattice parameters see the text. The energy barrier in the
superlattice is 177 meV and is indicated in the figure by an orizontal
line. The zero of the energy scale is the bottom of the GaAs conduction
band.}

\begin{table}
\setdec 0.000
\caption{First seven electron eigenvalues for 20 coupled quantum wells
(30/30 superlattice) with B=5T at different $L$ values. The cyclotron orbit
center value has been taken in the middle of the sample. The zero of the
energy scale is the bottom of the GaAs conduction band.}
\begin{tabular}{cccccccc}
 &\multicolumn{7}{c}{Electron energy levels (meV)}\\
\tableline
L=1 &\dec 106.84 &\dec 114.11 &\dec 121.26 &\dec 128.29 &\dec
135.15 &\dec 141.81 &\dec 148.18 \\
L=2 &\dec 104.13 &\dec 113.20 &\dec
119.32 &\dec 126.27 &\dec 134.13 &\dec 139.24 &\dec 145.28 \\
 L=3 &\dec
103.93 &\dec 113.02 &\dec 118.92 &\dec 125.69 &\dec 133.55 &\dec 138.45
&\dec 144.93 \\
\end{tabular}
\label{table1}
\end{table}

\begin{table}
\setdec 0.000
\caption{First seven superlattice eigenvalues for 20 coupled quantum wells
(30/30 superlattice) with $B||z$ and for the same superlattice potential
with an implemented EW. The magnetic field is B=50T and Landau level index
is n=0.The zero of the energy scale is the bottom of the GaAs conduction
band.}
\begin{tabular}{cccccccc}
 &\multicolumn{7}{c}{Electron energy levels (meV)}\\
\tableline
SL &\dec 97.97 &\dec 98.67 &\dec 99.83 &\dec 101.46 &\dec 103.52
&\dec 106.03 &\dec 108.95 \\ EW &\dec 96.23 &\dec 97.32 &\dec 98.51 &\dec
99.62 &\dec 102.12 &\dec 105.81 &\dec 107.41 \\
\end{tabular}
\label{table2}
\end{table}


\begin{references}
\bibitem[*]{note} Author to whom the correspondence should be addressed.
\bibitem{pa:ra}
We have used the following material parameters ($m_{o}$ is the free
electron mass, $x$ is the aluminum concentration):\\ band gap (1.5192 +
1.247 $x$) eV, electron effective mass (0.067 + 0.083 $x$) $m_{o}$, heavy
hole effective mass (0.45 + 0.31 $x$) $m_{o}$, light hole effective mass
(0.082 + 0.083 $x$) $m_{o}$, band offset $\Delta E_{c} = 0.62 \Delta E_{g}$
(where $\Delta E_{g}$ and $\Delta E_{c}$ are the band gap and conduction
band discontinuities between GaAs and Al$_{x}$Ga$_{1-x}$As).
\end{references}
\end{document}